\documentclass{article}
\usepackage{epsf}
\usepackage{amsmath}
\usepackage{amsfonts}
\usepackage{amssymb}

\newcommand{\bea}{\begin{eqnarray}}
\newcommand{\eea}{\end{eqnarray}}
\newcommand{\non}{\nonumber}

\begin{document}

\title{\textbf{Spin-$1$ Antiferromagnetic Heisenberg Chains in an External Staggered Field}}
\author{E. Ercolessi$^{(1)}$, G. Morandi$^{(1)}$, P. Pieri$^{(2)}$, and 
M. Roncaglia$^{(1)}$}
\date{\small  }

\maketitle
\vspace{-1cm}
\begin{center}
{\footnotesize (1) Dipartimento di Fisica, Universit\`a di Bologna, INFN and INFM,\\ V.le Berti Pichat 6/2, I-40127, Bologna, Italy\\
(2) Dipartimento di Matematica e Fisica and INFM, 
Universit\`{a} di Camerino\\[-2.5pt] V. Madonna delle Carceri, I-62032 Camerino, Italy}
\end{center}

\begin{abstract}
We present in this paper a nonlinear sigma-model analysis of a spin-$1$
antiferromagnetic Heisenberg chain in an external commensurate staggered
magnetic field. After rediscussing briefly and extending previous results
for the staggered magnetization curve, the core of the paper is a novel
calculation, at the tree level, of the Green functions of the model. 
We obtain precise results for the elementary excitation spectrum and in
particular for the spin gaps in the transverse and longitudinal channels. 
It is shown that, while the spectral weight in the transverse channel is
exhausted by a single magnon pole, in the longitudinal one, besides a magnon
pole a two-magnon continuum appears as well whose weight is a stedily
increasing function of the applied field, while the weight of the magnon
decreases correspondingly. The balance between the two is governed by a sum
rule that is derived and discussed. A detailed comparison with the present
experimental and numerical (DMRG) status of the art as well as with
previous analytical approaches is also made.
\end{abstract}


\section{Introduction}

In recent times there has arisen a great experimental \cite{1}-\cite{4} and theoretical
\cite{5}-\cite{7} interest in a new class of magnetic materials of the general
composition $R_{2}$BaNiO$_{5}$, with $R$ one of the magnetic rare-earth ions
(typically: $R=$Nd or $R=$Pr). These materials are obtained by substitution 
from the reference compound Y$_{2}$BaNiO$_{5}$, a highly one-dimensional 
compound with negligible interactions among the spin-$1$ Ni$^{2+}$ linear chains. For this reason Y$_{2}$BaNiO$_{5}$ is generally considered as an almost ideal example of an $S=1$ Haldane-gap system, with a spin gap of: $\Delta_{0}=0.41048(2)$ in units of the AFM intrachain exchange coupling \cite{8,9}.

As the magnetic $R^{3+}$ ions order antiferromagnetically below a certain
N\'{e}el temperature $T_{N}$, the $R_{2}$BaNiO$_{5}$'s have been modelled, to a
first approximation, as a set of $S=1$ chains with a negligible interchain
coupling (as compared with the intrachain one), acted upon by an effective
commensurate staggered field \cite{2} roughly proportional to the sublattice
magnetization of the $R^{3+}$ lattice, and hence increasing when the
temperature is decrease below $T_{N}$.

The above experimental scenario has motivated a renewal of theoretical
activity on the model, that is nonetheless already quite interesting ``per
se'', of an integer-spin AFM Heisenberg chain coupled to an external
commensurate staggered field, that can be described by the model Hamiltonian:
\begin{equation}
\mathcal{H}=\sum\limits_{i}\{J\mathbf{S}_{i}\cdot\mathbf{S}_{i+1}%
+(-1)^{i}\mathbf{H}_{s}\cdot\mathbf{S}_{i}\}
\end{equation}
where: $\mathbf{S}_{i}^{2}=S(S+1)$ with $S$ an integer (we set $\hbar=1$
henceforth, and take $S=1$ for the $Ni^{2+}$ chains), $J>0$ and $\mathbf{H}%
_{s}$ is the external staggered field in appropriate units (see, e.g., Ref.
\cite{6} for details).

An extensive DMRG study of the model of Eq.$(1)$ has been performed in
Ref.\cite{6}, where very accurate results were reported for the staggered
magnetization curve, the spin gaps, the static correlation functions and the
correlation lengths in both the longitudinal and transverse (with respect to  the direction of the field) channels. 
The authors of Ref.\cite{5} made instead an analytic study of the model beginning with the familiar mapping \cite{10,11} of
the Hamiltonian $(1)$ onto a nonlinear sigma-model (NL$\sigma$M). What they
discussed very accurately was actually a related and somewhat more
phenomenological model in which the strict NL$\sigma$M constraint is
softened, then replacing the original NL$\sigma$M with a theory of the
Ginzburg-Landau type parametrized by an adequate set of adjustable parameters
(see Ref.\cite{5} for details).

One of the purposes of the present paper is to investigate carefully what are
the resulting similarities and/or differences when the NL$\sigma$M constraint
is not softened but enforced consistently at each level of approximation. We
will report here only results at the tree-level of a loop expansion \cite{12},
i.e. essentially at the the mean-field (MFT) level, of the partition
function of the model, supplementing them however with a stability analysis,
and deferring a systematic evaluation of higher loop corrections, that are
considerably more involved, to a forthcoming paper \cite{13}.

Our second purpose is to assess the validity in the present context of the 
so-called single-mode approximation (SMA) that did prove beyond doubt its validity previously but in rather different contexts \cite{14}. 
We will provide here the explicit proof
of the fact that indeed the SMA is not applicable to discuss the elementary
excitation spectrum in the longitudinal channel, a claim that we had already
put forward some time ago \cite{7}, without providing however there an explicit proof.

The paper is organized as follows. In Sect.$2$ we state the essentials of the
general formalism and derive the saddle-point approximation in the presence of
a general external source field. This is needed in order to set up the
consistent scheme of calculation of the propagators at the mean-field level 
that is reported in Sect.$3$. 
In Sect.$4$ we study the analytic structure of the
propagators, and notably of the longitudinal one, at the physical
saddle-point, i.e. when the source field becomes the staggered static field of
Eq. $(1)$. The final Sect.$5$ is devoted to a discussion of our results and to
a detailed comparison with previous theoretical approaches, as well as to a
discussion of some as yet unsolved problems that are posed by the experimental
scenario that has been outlined at the beginning. A useful sum rule for the
propagators of the NL$\sigma$M field is derived in the Appendix.

\section{Saddle-point approximation for a general\\ source field}

Under the Haldane mapping \cite{11}: $\mathbf{S}_{i}\approx(-1)^{i}%
S\mathbf{n}_{i}+\mathbf{l}_{i}$, $\mathbf{n}_{i}^{2}=1$, where $\mathbf{n}%
_{i}$ represents the slowly-varying local staggered magnetization and
$\mathbf{l}_{i\text{ }}$is the local generator of angular momentum, the Zeeman
term of Eq.$(1)$ becomes : $\sum_{i}(-1)^{1}\mathbf{H}_{s}\cdot\mathbf{S}%
_{i}\approx S\sum_{i}\mathbf{n}_{i}\cdot\mathbf{H}_{s}+\sum_{i}(-1)^{1}%
\mathbf{l}_{i}\cdot\mathbf{H}_{s}$. In the continuum limit the second term
becomes a total derivative that can be neglected if we adopt periodic boundary
conditions on the chain.

Going then to the continuum limit, integrating out the fluctuation field
$\mathbf{l}$ and implementing the NL$\sigma$M constraint $\mathbf{n}%
^{2}(\mathbf{x)}=1$ ($\mathbf{x=(}x,\tau)$ with $\tau$ the Euclidean time)
with the aid of a Lagrange multiplier $\lambda=\lambda(\mathbf{x})$ , we
obtain the partition function: $\frak{Z=}Tr\{\exp[-\beta\mathcal{H}]\}$ of the
model in the continuum limit as the path-integral:%

\begin{equation}
\frak{Z}=\int[\mathcal{D}\mathbf{n}]\left[\frac{\mathcal{D}\lambda}{2\pi}
\right]\exp(-S_{eff})
\end{equation}
where the effective action $S_{eff}$ is given by:%
\begin{equation}
S_{eff}=\int d\mathbf{x}\left\{\mathcal{L}_{E}(\mathbf{x})-S\mathbf{H}_{s}
\cdot\mathbf{n}(\mathbf{x})-i\lambda(\mathbf{x)(n}^{2}(x)-1)\right\}
\end{equation}
where: \ $\int d\mathbf{x}$ $=\int dx\int_{0}^{\beta}d\tau$ and the Euclidean
Lagrangian is given by:
\begin{equation}
\mathcal{L}_{E}(\mathbf{x})=\frac{1}{2gc}\left(c^{2}|\partial_{x}\mathbf{n}
|^{2}+|\partial_{\tau}\mathbf{n}|^{2}\right)
\end{equation}
and the NL$\sigma$M mapping predicts: $g=2/S$ for the coupling constant and:
$c=2JSa$ (with $a$ the lattice constant) for the spin-wave velocity.

Now we promote $\frak{Z}$ to a generating functional $\frak{Z}[\mathbf{J}]$ by
replacing $S_{eff}$ with:
\begin{equation}
S[\mathbf{J}]=\int d\mathbf{x}\left\{\mathcal{L}_{E}(\mathbf{x})-S\mathbf{J}%
(\mathbf{x})\cdot\mathbf{n}(\mathbf{x})-i\lambda(\mathbf{x)(n}^{2}(x)-1)\right\}
\end{equation}
and we will set: $\mathbf{J=H}_{s}$ only at the end of the calculations.

Altogether (after an integration by parts):
\begin{equation}
S[\mathbf{J}]=\frac{1}{2}\int d\mathbf{x}d\mathbf{x}^{\prime}\,\mathbf{n}%
(\mathbf{x})\cdot G_{0}^{-1}(\mathbf{x},\mathbf{x}^{\prime})\mathbf{n}%
(\mathbf{x}^{\prime})-S\int d\mathbf{x\,J(x)\cdot n(x)}+i\int d\mathbf{x}\,%
\lambda(\mathbf{x})
\end{equation}
and $G_{0}$ solves, with the appropriate (Matsubara-Bose) boundary conditions
the equation:%
\begin{equation}
-\frac{1}{gc}(c^{2}\partial_{x}^{2}+\partial_{\tau}^{2}+2igc\lambda
(\mathbf{x}))G_{0}(\mathbf{x,x}^{\prime})=\delta^{(2)}(\mathbf{x-x}^{\prime})
\end{equation}

Performing now the linear shift: $\mathbf{n(x)}=\mathbf{n}^{\prime
}(x)+\mathbf{a(x)}$, with:%
\begin{equation}
\mathbf{a(x)=}S\int d\mathbf{x}^{\prime}G_{0}(\mathbf{x,x}^{\prime
})\mathbf{J(x}^{\prime})
\end{equation}
we obtain:
\bea
S[\mathbf{J}]&=&\frac{1}{2}\int d\mathbf{x}d\mathbf{x}^{\prime}\mathbf{n}^{\prime}(\mathbf{x})\cdot G_{0}^{-1}(\mathbf{x},\mathbf{x}^{\prime})\mathbf{n}^{\prime}(\mathbf{x}^{\prime})+i\int d\mathbf{x}\lambda (\mathbf{x})\non\\ &&  -\frac{1}{2}S^{2}\int d\mathbf{x}d\mathbf{x}^{\prime}\mathbf{J(x)\cdot}
 G_{0}(\mathbf{x,x}^{\prime})\mathbf{J(x}^{\prime})
\eea

Now we can integrate out the field $\mathbf{n}^{\prime}$, obtaining:%
\begin{equation}
\frak{Z}[\mathbf{J]}\propto \int\left[\frac{\mathcal{D}\lambda}{2\pi}%
\right]\exp(-S[\lambda;\mathbf{J}])
\end{equation}
where:%
\begin{equation}
S[\lambda;\mathbf{J}]=\frac{3}{2}Tr\{\ln(G_{0}^{-1})\}+i\int d\mathbf{x}%
\lambda(\mathbf{x})-\frac{1}{2}S^{2}\int d\mathbf{x}d\mathbf{x}^{\prime
}\mathbf{J(x)\cdot}G_{0}(\mathbf{x,x}^{\prime})\mathbf{J(x}^{\prime})
\end{equation}

We analyze now what are the general features of a saddle-point approximation
made in the presence of an arbitrary (space-time dependent) source field
$\mathbf{J(x)}$. We will also evaluate here the propagators at the mean-field 
level. The saddle point will be determined by the equation:%
\begin{equation}
\left(\frac{\delta S[\lambda;\mathbf{J}]}{\delta\lambda(\mathbf{x}%
)}\right)_{\mathbf{J}}=0
\end{equation}
where $(..)_{\mathbf{J}}$ means that we (functionally) differentiate while
keeping $\mathbf{J}$ constant. As:%
\begin{equation}
\frac{\delta G_{0}(\mathbf{x}^{\prime}\mathbf{,x}^{\prime\prime})}%
{\delta\lambda(\mathbf{x})}=2i G_{0}(\mathbf{x}^{\prime},\mathbf{x})G_{0}(\mathbf{x,x}^{\prime\prime})
\end{equation}
and:
\begin{equation}
i\frac{\delta S}{\delta\lambda(\mathbf{x})}=3 G_{0}(\mathbf{x,x})+S^{2}\int
d\mathbf{y}d\mathbf{y}^{\prime} G_{0}(\mathbf{y,x}) G_{0}(\mathbf{x,y}%
^{\prime})[\mathbf{J(y})\cdot\mathbf{J(y}^{\prime})]-1
\end{equation}
we find the saddle-point equation in the form:%

\begin{equation}
3 G_{0}(\mathbf{x,x})+S^{2}\int d\mathbf{y}d\mathbf{y}^{\prime} G_{0}
(\mathbf{y,x}) G_{0}(\mathbf{x,y}^{\prime})[\mathbf{J(y})\cdot\mathbf{J(y}%
^{\prime})]=1
\end{equation}
Eq.(15) will determine then a space-time dependent saddle-point that will be a
functional of $\mathbf{J}$ as well: $\lambda=\lambda^{\ast}[\mathbf{x;J}]$.

In Mean-Field Theory (MFT) one approximates $\frak{Z}[\mathbf{J}]$ as:
\begin{equation}
\frak{Z}[\mathbf{J}]\approx\exp(-S[\lambda^{\ast};\mathbf{J}])
\end{equation}
and hence the connected two-point propagators:%
\begin{equation}
G_{c}^{\alpha\beta}(\mathbf{x,x}^{\prime})=S^{2}\left\{\left\langle T(n^{\alpha
}(\mathbf{x})n^{\beta}(\mathbf{x}^{\prime})\right\rangle -\left\langle
n^{\alpha}(\mathbf{x})\right\rangle \left\langle n^{\beta}(\mathbf{x}^{\prime
})\right\rangle \right\}=\frac{\delta^{2}\ln\frak{Z}[\mathbf{J}]}{\delta J^{\alpha
}(\mathbf{x)}\delta J^{\beta}(\mathbf{x}^{\prime})}%
\end{equation}
will be given by:%
\begin{equation}
G_{c}^{\alpha\beta}(\mathbf{x,x}^{\prime})\approx-\frac{\delta^{2}%
S[\lambda^{\ast};\mathbf{J}]}{\delta J^{\alpha}(\mathbf{x)}\delta J^{\beta
}(\mathbf{x}^{\prime})}%
\end{equation}
where now, when taking functional derivatives, one has to consider
not only the explicit dependence of $S$ on $\mathbf{J}$ but also the implicit
one through $\lambda^{\ast}$. We find then:%
\begin{equation}
\frac{\delta S}{\delta J^{\alpha}(\mathbf{x})}=\int d\mathbf{y}\left(
\frac{\delta S}{\delta\lambda(\mathbf{y})}\right)_{\mathbf{J}}\frac
{\delta\lambda(\mathbf{y})}{\delta J^{\alpha}(\mathbf{x})}+\left(
\frac{\delta S}{\delta J^{\alpha}(\mathbf{x})}\right)_{\mathbf{\lambda}}%
\end{equation}
But the first term on the r.h.s. vanishes identically at the saddle point,
and so:
\begin{equation}
\frac{\delta S}{\delta J^{\alpha}(\mathbf{x})}=\left(\frac{\delta
S}{\delta J^{\alpha}(\mathbf{x})}\right)_{\mathbf{\lambda}}=-\frac{1}%
{2}S^{2}\int d\mathbf{y}[G_{0}(\mathbf{x,y})+ G_{0}(\mathbf{y,x})]J^{\alpha
}(\mathbf{y)}%
\end{equation}
which provides also the mean-field equation for the (staggered) ``magnetization'' induced by the source field $\mathbf{J}$.

Proceeding one step further we find eventually:
\bea
G_{c}^{\alpha\beta}(\mathbf{x,x}^{\prime})&=&\frac{1}{2}S^{2}[G_{0}
(\mathbf{x,x}^{\prime})+(\mathbf{x\longleftrightarrow x}^{\prime}%
)]\delta^{\alpha\beta}\non\\
&& \hspace*{-1.5cm}+S^{2}\int d\mathbf{y}d\mathbf{y}^{\prime}%
[G_{0}(\mathbf{x,y)} G_{0}(\mathbf{y,y}^{\prime})+(\mathbf{x\longleftrightarrow y}^{\prime})]J^{\alpha}(\mathbf{y})\left(i\frac{\delta\lambda(\mathbf{y}^{\prime})}{\delta J^{\beta
}(\mathbf{x}^{\prime})}\right)
\eea

This is the general structure of the mean-field propagators for a general source
field $\mathbf{J}$.

The functional derivative of $\lambda$ on the r.h.s. of Eq.(21) is determined by:
\begin{equation}
0=\frac{\delta}{\delta J^{\alpha}}\left(\frac{\delta S}{\delta\lambda
}\right)_{\mathbf{J}}=\left(\frac{\delta^{2}S}{\delta\lambda
\delta\lambda^{\prime}}\right)_{\mathbf{J}}\frac{\delta\lambda^{\prime}%
}{\delta J^{\alpha}}+\left(\frac{\delta}{\delta J^{\alpha}}\left(
\frac{\delta S[\lambda;\mathbf{J}]}{\delta\lambda}\right)_{\mathbf{J}}\right)_{\lambda}%
\end{equation}
where the second term represents the variation of $(\delta S/\delta
\lambda)_{\mathbf{J}}$ w.r.t. its explicit dependence on $\mathbf{J}$. As all
the quantities in brackets have to be evaluated at the saddle point, this
equation becomes an inhomogeneous linear integral equation for $\delta
\lambda/\delta J^{\alpha}$ whose kernel is:
\bea 
H(\mathbf{x,x}^{\prime})&=&\left(\frac{\delta^{2}S}{\delta\lambda
(\mathbf{x})\delta\lambda(\mathbf{x}^{\prime})}\right)_{\mathbf{J}
}|_{\lambda=\lambda^{\ast}}\non\\ 
&=& 6\Gamma(\mathbf{x,x}^{\prime})+4S^{2}
G_{0}(\mathbf{x}^{\prime}\mathbf{,x})\int d\mathbf{y}d\mathbf{y}^{\prime
} G_{0}(\mathbf{y,x}^{\prime}) G_{0}(\mathbf{x,y}^{\prime})[\mathbf{J(y}
)\cdot\mathbf{J(y}^{\prime})]\non\\
\eea
where $\Gamma$ is the ``polarization bubble'':
\begin{equation}
\Gamma(\mathbf{x,x}^{\prime})= G_{0}(\mathbf{x,x}^{\prime}) G_{0}
(\mathbf{x}^{\prime}\mathbf{,x})
\end{equation}

The integral equation for $\delta\lambda/\delta J^{\alpha}(\mathbf{x})$ reads then:

\begin{equation}
\int d\mathbf{y}H(\mathbf{x,y})\left(i\frac{\delta\lambda(\mathbf{y}%
)}{\delta J^{\alpha}(\mathbf{x}^{\prime})}\right)=-2S^{2} G_{0}
(\mathbf{x,x}^{\prime})\int d\mathbf{y} G_{0}(\mathbf{x,y})J^{\alpha
}(\mathbf{y})
\end{equation}

In the next Section we will specialize the results obtained here to the
physical case $\mathbf{J}=const.=\mathbf{H}_{s}$, which will determine the
physically relevant saddle point.

\section{Results at the physical saddle point}

When $\ \mathbf{J}=const.=\mathbf{H}_{s}$, the associated saddle point will
correspond also to $\lambda=const.$ \ 
Setting then: $-2igc\lambda=const.=c^{2}%
\xi^{-2}$, translational invariance will be restored and the following results
can be easily derived \cite{7}:
\begin{enumerate}
\item  The saddle-point condition is:
\begin{equation}
3 G_{0}(\mathbf{0})+S^{2}\mathbf{H}_{s}^{2}[\widetilde{G_{0}}(\mathbf{0}%
)]^{2}=1
\end{equation}
where $\widetilde{G_{0}}(\mathbf{0})=\widetilde{G_{0}}(\mathbf{q=0})$
($\mathbf{q}=(q,\Omega_{n}=2\pi n/\beta$) and $\widetilde{G_{0}}(\mathbf{q}%
)$, the Fourier transform of $G_{0}(\mathbf{x})$, is given by:%
\begin{equation}
\widetilde{G_{0}}(\mathbf{q})=\frac{gc}{\Omega_{n}^{2}+c^{2}(q^{2}+\xi^{-2})}%
\end{equation}
whence: $\widetilde{G_{0}}(0)=g\xi^{2}/c.$ Explicitly, at $T=0$ \cite{7}:
\begin{equation}
\frac{3g}{2\pi}\ln\left\{\Lambda\xi+\sqrt{1+(\Lambda\xi)^{2}}\right\}=
1-\left(\frac{Sg}{c}\right)^{2}\mathbf{H}_{s}^{2}\xi^{4}
\end{equation}
and the cutoff \ can disposed of by fitting it to the zero-field gap
$\Delta_{0}=c\xi^{-1}$ that is know from\ the DMRG studies. Eq.($28$) will
determine then the field dependence of the correlation length $\xi$, and it is
clear that $\xi$ will depend quadratically on the field.

\item  The magnetization is given by:
\begin{equation}
\mathbf{m}_{s}=S\left\langle \mathbf{n}_{s}\right\rangle =S^{2}\widetilde
{\Delta}(\mathbf{0})\mathbf{H}_{s}=\frac{gS^{2}\xi^{2}}{c}\mathbf{H}_{s}%
\end{equation}
Comparison with the DMRG data of Ref.\cite{6} shows a slight overestimate of the
values of the magnetization for small fields, but the agreement becomes better
and better as the field increases (see Fig.$1$).

\begin{figure}[t]
\begin{center}
\hspace*{.1cm}
\epsfxsize=7cm \epsffile{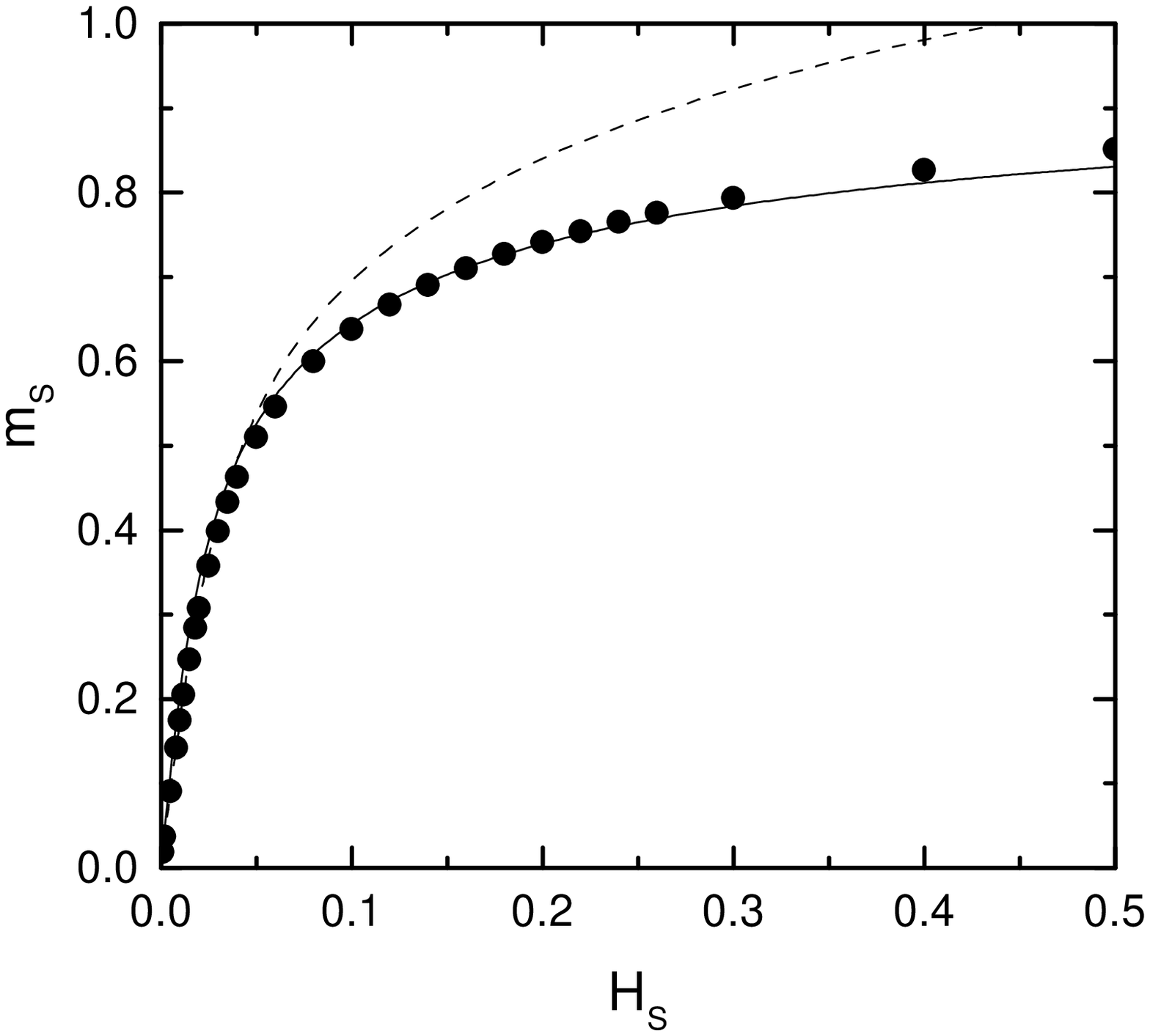}
\vspace{.4cm}
\caption[]{\footnotesize 
Staggered magnetization for $S=1$ as a function of the staggered field:\\
Dots: DMRG results of Ref.\cite{6}. 
Full line: our results. 
Dashed line: results of Ref.\cite{5}.}
\end{center}
\label{fig3}
\end{figure}

Note that, in view of this equation, the saddle-point condition can be written
also as:%
\begin{equation}
3 G_{0}(\mathbf{0})+\left(\frac{m_{s}}{S}\right)^{2}=1
\end{equation}

\item  While the transverse susceptibility is given by: $\chi^{T}=m_{s}%
/H_{s}=gS^{2}\xi^{2}/c$, the longitudinal one is given by the full derivative:
$\chi^{L}=dm_{s}/dH_{s}$, i.e. by:
\begin{equation}
\chi^{L}=\chi^{T}\left\{1+(2H_{s}/\xi)\frac{d\xi}{dH_{s}}\right\}=
\chi^{T}\left\{1+\frac{d(\ln(\xi^{2}))}{d(\ln H_{s})}\right\}
\end{equation}
The derivative on the r.h.s. can be obtained by differentiating the
saddle-point equation w.r.t. $H_{s}$, and the explicit expression for
$\chi^{L}$ is:
\begin{equation}
\chi^{L}=\chi^{T}\left\{1+\frac{2\pi}{3g}\left(\frac{2gS\xi^{2}H_{s}}{c}\right)^{2}\frac
{\sqrt{1+(\Lambda\xi)^{2}}}{\Lambda\xi}\right\}^{-1}
\end{equation}
(clearly exhibiting: $\chi^{L}<\chi^{T}$ always, the two coinciding only when
$H_{s}=0$). 
In the limit $\Lambda\xi\gg1$, Eq.$(32)$ reduces then to:
\begin{equation}
\chi^{L}=\chi^{T}\frac{1}{1+\frac{2\pi}{3g}(2gS\xi^{2}H_{s}/c)^{2}}%
\end{equation}

\item  In the translationally-invariant case, Eq.(25) for $\delta
\lambda/\delta J^{\alpha}$ can be rewritten, using the equation for the
magnetization, as:%
\begin{equation}
\int d\mathbf{y}H(\mathbf{x-y})\left(i\frac{\delta\lambda(\mathbf{y}%
)}{\delta J^{\alpha}(\mathbf{x}^{\prime})}\right)=-2m_{s}^{\alpha}%
G_{0}(\mathbf{x-x}^{\prime})
\end{equation}

Making then the ``Ansatz'':
\begin{equation}
i\frac{\delta\lambda(\mathbf{x})}{\delta J^{\alpha}(\mathbf{x}^{\prime}%
)}=m_{s}^{\alpha}X(\mathbf{x}-\mathbf{x}^{\prime})
\end{equation}
(Hence: $\delta\lambda/\delta J^{\alpha}=0$ in the directions orthogonal to
the field) Eq.(34) reduces to the following equation for $X$:%
\begin{equation}
\int d\mathbf{y}H(\mathbf{x-y})X(\mathbf{y}-\mathbf{x}^{\prime})=-2 G_{0}
(\mathbf{x}-\mathbf{x}^{\prime})
\end{equation}
that can be solved by Fourier transforming it, thus yielding:%
\begin{equation}
\widetilde{X}(\mathbf{q})=-2\frac{\widetilde{G_{0}}(\mathbf{q})}%
{\widetilde{H}(\mathbf{q})}%
\end{equation}
where, now:
\begin{equation}
\widetilde{H}(\mathbf{q})=6\widetilde{\Gamma}(\mathbf{q})+4\left(\frac{m_{s}}%
{S}\right)^{2}\widetilde{G_{0}}(\mathbf{q})
\end{equation}

\item  The longitudinal connected propagator is given by:%
\begin{equation}
G_{c}^{L}(\mathbf{x}-\mathbf{x}^{\prime})=S^{2}G_{0}(\mathbf{x}%
-\mathbf{x}^{\prime})-2m_{s}^{2}\int d\mathbf{y}G_{0}(\mathbf{x}%
-\mathbf{y})X(\mathbf{y}-\mathbf{x}^{\prime})
\end{equation}
while:

\item  The transverse propagator (that has no nonconnected parts) is simply
given by:%
\begin{equation}
G_{c}^{T}(\mathbf{x}-\mathbf{x}^{\prime})=S^{2} G_{0}(\mathbf{x}%
-\mathbf{x}^{\prime})
\end{equation}
Going to Fourier space:%
\begin{equation}
\widetilde{G}_{c}^{T}(\mathbf{q})=S^{2}\widetilde{G_{0}}(\mathbf{q})
\end{equation}
which is (see Eq.($27$)) a free boson propagator that, when analytically
continued to the real axis, has simple poles at: $\omega=\pm\varepsilon(q)$
with:
\begin{equation}
\varepsilon(q)=\sqrt{c^{2}q^{2}+\Delta_{0}^{2}},\text{ }\Delta_{0}=c\xi^{-1}%
\end{equation}
Notice also that:%
\begin{equation}
\chi^{T}=\widetilde{G}_{c}^{T}(\mathbf{0})
\end{equation}
\end{enumerate}

Explicitly, in Fourier space, one finds: 
$\widetilde{G_{c}^{L}}(\mathbf{q}) = S^2\widetilde{G_{0}}(\mathbf{q})-2m_{s}^{2}
\widetilde{G_{0}}(\mathbf{q})\widetilde{X}(\mathbf{q})$ and, with some algebra:
\begin{equation}
\widetilde{G}_{c}^{L}(\mathbf{q})=S^{2}\widetilde{G_{0}}(\mathbf{q}%
)\frac{3\widetilde{\Gamma}(\mathbf{q})}{3\widetilde{\Gamma}(\mathbf{q}%
)+2(m_{s}/S)^{2}\widetilde{G_{0}}(\mathbf{q})}\equiv\widetilde{G}%
_{c}^{T}(\mathbf{q})\frac{3\widetilde{\Gamma}(\mathbf{q})}{3\widetilde{\Gamma
}(\mathbf{q})+2(m_{s}/S)^{2}\widetilde{G_{0}}(\mathbf{q})}%
\end{equation}
$\widetilde{\Gamma}(\mathbf{q})$ is the convolution of two $\widetilde{G_{0}}$'s that can be evaluated explicitly as:
\bea
\widetilde{\Gamma}(\mathbf{q})&=&\frac{1}{2}(gc)^{2}\int\frac{dk}{2\pi}
\coth(\beta\varepsilon(k)/2)\frac{1}{\varepsilon(k)\varepsilon(k+q)}\times\non\\
&&\qquad \left\{\frac{\varepsilon(k+q)+\varepsilon(k)}{\Omega_{n}^{2}%
+(\varepsilon(k+q)+\varepsilon(k))^{2}}+\frac{\varepsilon(k+q)-\varepsilon
(k)}{\Omega_{n}^{2}+(\varepsilon(k+q)-\varepsilon(k))^{2}}\right\}
\eea

The analytic continuation in frequency ($\Omega_{n}\mapsto iz$) has a
branch-cut along the entire real axis for all $\beta<+\infty$. The
discontinuity across the branch-cut vanishes however exponentially with
temperature in the range: $-2\Delta_{0}(q)<\operatorname{Re}(z)<+2\Delta
_{0}(q)$, where: $\Delta_{0}(q)=c\sqrt{(q/2)^{2}+\xi^{-2}}=\varepsilon(q/2)$
and, right at $T=0$, the branch cut extends only from $-\infty$ to
\ $-2\Delta_{0}(q)$ and from $+2\Delta_{0}(q)$ to $+\infty$. The reason for
this is that the second term in curly brackets vanishes exponentially for
$\beta\mapsto\infty$. The same term vanishes also, irrespective of
temperature, for $q=0$. For this reason we expect also $\widetilde{\Gamma
}(\mathbf{q})$ to be an essentially positive quantity at all temperatures.
According \ also to Eq.$(27)$, this implies that $\widetilde{H}(\mathbf{q})$
is a positive-definite quantity, and hence that $H(\mathbf{x}-\mathbf{x}%
^{\prime})$ is a positive-definite kernel.

Now, the kernel $H$ determines actually also the stability of the saddle
point. Indeed, by expanding the action (11) quadratically around
the (physical) saddle point in terms of: 
$\delta\lambda(\mathbf{x})=\lambda(\mathbf{x})-\lambda^{\ast}$, we obtain
\begin{equation}
S[\lambda]=S[\lambda^{\ast}]+\frac{1}{2}\int d\mathbf{x}d\mathbf{x}^{\prime
}\delta\lambda(\mathbf{x})H(\mathbf{x}-\mathbf{x}^{\prime})\delta
\lambda(\mathbf{x}^{\prime})
\end{equation}
and the positivity of $H$ will then guarantee that the physical saddle point
is indeed differentially stable, i.e. a local minimum.

We analyze now the asymptotic behaviour of the longitudinal propagator for
both large and small values of $|\mathbf{q}|$.

While one sees immediately that: $\widetilde{\Delta}(\mathbf{q})\approx
|\mathbf{q}|^{-2}$ for large $|\mathbf{q}|$, in the same limit \cite{15}:
$\widetilde{\Gamma}(\mathbf{q})\approx\ln(|\mathbf{q}|^{2})/|\mathbf{q}|^{2}$, instead. Therefore: $\widetilde{G_{c}^{L}}(\mathbf{q})$ $\approx$
$\widetilde{G_{c}^{T}}(\mathbf{q})$ for large $|\mathbf{q}|$, and the (very)
short-distance behaviour of the two propagators is the same. This implies:%
\begin{equation}
\underset{\mathbf{x}\mapsto0}{\lim}G_{c}^{L}(\mathbf{x})=\underset
{\mathbf{x}\mapsto0}{\lim}G_{c}^{T}(\mathbf{x})=S^{2}G_{0}(\mathbf{0})
\end{equation}
Recalling that, in the translationally-invariant case, the full longitudinal
propagator $G^{L}$ is related to the connected one by: $G^{L}=G_{c}^{L}%
+m_{s}^{2}$, we see that, as a consequence: $3G_{0}(\mathbf{0})+(m_{s}%
/S)^{2}\equiv S^{-2}\{2G^{T}(\mathbf{0})+G^{L}(\mathbf{0})\}$, and that
therefore the saddle-point condition can be again read simply as one
implementing the constraint on the average, i.e. as:%
\begin{equation}
\left\langle \mathbf{n}^{2}\right\rangle =1
\end{equation}

At the opposite end, when $\mathbf{q\mapsto0}$:%

\begin{equation}
\widetilde{G}_{c}^{L}(\mathbf{0})=\frac{3S^{2}\widetilde{\Gamma
}(\mathbf{0})\widetilde{G_{0}}(\mathbf{0})}{3\widetilde{\Gamma}%
(\mathbf{0})+2(m_{s}/S)^{2}\widetilde{G_{0}}(\mathbf{0})}%
\end{equation}
that is markedly different from $\widetilde{G_{c}^{T}}(\mathbf{0})$,
coinciding with the latter only for $H_{s}\mapsto0$. The same will be true for
the small-momentum behaviour of $\widetilde{G_{c}^{L}}(\mathbf{q})$. The
long-distance behaviours of the two propagators will be then definitely
different for $H_{s}\neq0$, and so we expect quite different asymptotic
behaviours at infinity, i.e. quite different correlation lengths \cite{6}.

On top of that, the equation (cfr, Eq.(43)): $\chi^L=\widetilde{G_{c}^{L}}(\mathbf{0})$ provides us also with
an independent expression for the longitudinal susceptibility,
explicitly showing how it results from both one- and two-magnon contributions
(the latter coming from the ``polarization bubble'' $\widetilde{\Gamma
}(\mathbf{q)}$).

In explicit terms, we have, at $T=0$:%
\begin{equation}
\widetilde{\Gamma}(\mathbf{0})=\frac{(gc)^{2}}{8\pi}\int dk\frac
{1}{\varepsilon(k)^{3}}=\frac{g^{2}}{4\pi c}\int\limits_{0}^{\infty}%
dk(k^{2}+\xi^{-2})^{-\frac{3}{2}}=\frac{(g\xi)^{2}}{4\pi c}%
\end{equation}
Inserting this expression (together with the known value of $\widetilde
{G_{0}}(\mathbf{0})$) into the equation for $\widetilde{G}_{c}%
^{L}(\mathbf{0})$ yields back precisely Eq.$(33)$ that had been
obtained by letting $\Lambda\mapsto\infty$ in the previous equation for
the longitudinal susceptibility wherever this did not lead to divergent
results, which is precisely the attitude that has been taken in the present
calculation. All this shows that (at least) there are no inconsistencies in
the MFT approach that has been adopted here.

In the next Section we will discuss in detail the structure of the analytic 
continuation of the propagators to the complex frequency plane and to the real axis.

\section{The analytic structure of the propagators}

At the present level of approximation the transverse propagator is just a
free-boson propagator. Analytic continuation is straightforward, leading to:
\begin{equation}
\widetilde{G_{c}^{T}}(q,z)=\frac{gcS^{2}}{\varepsilon^{2}(q)-z^{2}}%
\end{equation}
($\varepsilon(q)=c\sqrt{q^{2}+\xi^{-2}}$). Going to the real axis from above:
$z=\omega+i\eta,\eta\mapsto0^{+}$:%
\begin{equation}
\operatorname{Im}\widetilde{G_{c}^{T}}(q,\omega)=\frac{\pi gcS^{2}
}{2\varepsilon(q)}\left\{\delta(\omega-\varepsilon(q))-
\delta(\omega+\varepsilon(q))\right\}
\end{equation}
The spectral weight function is then fully exhausted by single poles at
$\omega=\pm\varepsilon(q)$ which is the structure required \cite{5,14} for the
applicability of the SMA. The relation, which is a direct consequence \cite{5}
of the SMA: $\chi_{T}=Sgc/(\Delta_{T})^{2}$, with: $\Delta_{T}=\Delta
_{0}=c\xi^{-1}$, is obeyed exactly, at this level of approximation, in the
transverse channel.

Let's turn now to the longitudinal propagator, and let's begin by looking at
the polarization bubble $\widetilde{\Gamma}(\mathbf{q})=\widetilde{\Gamma
}(q,\Omega_{n})$. We shall consider for simplicity only the $T=0$ 
limit in which the second term on the r.h.s. of Eq.$(45)$ can be neglected.
Then:
\begin{equation}
\widetilde{\Gamma}(\mathbf{q})=\frac{1}{2}(gc)^{2}\int\frac{dk}{2\pi}\,\frac
{1}{\varepsilon(k)\varepsilon(k+q)}\,\frac{\varepsilon(k+q)+\varepsilon
(k)}{\Omega_{n}^{2}+(\varepsilon(k+q)+\varepsilon(k))^{2}}%
\end{equation}
that we write for short as:%
\begin{equation}
\widetilde{\Gamma}(\mathbf{q})=\frac{1}{2}(gc)^{2}\int\frac{dk}{2\pi}%
\,\frac{A(k,q)}{\Omega_{n}^{2}+E(k,q)^{2}}%
\end{equation}
where:%
\begin{equation}
E(k,q)=\varepsilon(k+q)+\varepsilon(k)
\end{equation}
and:%
\begin{equation}
A(k,q)=\frac{E(k,q)}{\varepsilon(k)\varepsilon(k+q)}%
\end{equation}
Note that :%
\begin{equation}
E(k,q)=E(-k-q,q)
\end{equation}
and the same will hold true for $A(k,q)$.

Performing now the analytic continuation and going to the real axis from
above:%
\begin{equation}
\widetilde{\Gamma}(q,\omega)=\Gamma_{1}(q,\omega)+i\Gamma_{2}(q,\omega)
\end{equation}
where:%
\begin{equation}
\Gamma_{1}(q,\omega)=\frac{1}{2}(gc)^{2}\int\frac{dk}{2\pi}\mathcal{P}%
\left\{\frac{A(k,q)}{E^{2}(k,q)-\omega^{2}}\right\}
\end{equation}
(``$\mathcal{P}$'' standing for Cauchy principal part) and:%
\begin{equation}
\Gamma_{2}(q,\omega)=\frac{1}{4}(gc)^{2}\int dk\frac{1}{\epsilon
(k)\epsilon(k+q)}[\delta(\omega-E(k,q))-\delta(\omega+E(k,q))]
\end{equation}
that is odd in $\omega$ (which implies that  $\Gamma_{1}(q,\omega)$ will be
an even function of $\omega$) and positive for positive $\omega$.

To evaluate $\Gamma_{2}$ explicitly it will be enough to consider the positive
frequency part. Notice that, for any fixed $q$: $\min\{E(k,q)\}=2\Delta
_{0}(q)$ ($\Delta_{0}(q)=\varepsilon(q/2)$, $\Delta_{0}(0)=\Delta_{0}$), and therefore: $\Gamma_{2}(q,\omega)=0$ for $|\omega|<2\Delta_{0}(q)$, as we know already.
Otherwise, it is easy to see graphically that the equation: $\omega=E(k,q)$
has two solutions at $k=k^{\ast}(q,\omega)=k_0-q/2$ and at 
$k=-k^{\ast}-q=k_0-q/2$, where: $k_{0}=(\omega/2c)\sqrt{(\omega^{2}%
-4\Delta_{0}^{2}(q))/(\omega^{2}-c^{2}q^{2})}$.

Then we obtain easily, as: $\varepsilon(k)\varepsilon(k+q)(\partial
E(k,q)/\partial k)=c^{2}[k\varepsilon(k+q)+$\break $(k+q)\varepsilon(k)]$:
\begin{equation}
\Gamma_{2}(q,\omega)=\frac{g^{2}}{4C(q)}\theta(\omega^{2}-4\Delta_{0}%
^{2}(q))sgn(\omega)
\end{equation}
where: $C(q)=|k^{\ast}\varepsilon(k^{\ast}+q)+(k^{\ast}+q)\varepsilon(k^{\ast
})|$ and: $k^{\ast}=k^{\ast}(q,|\omega|)$. An explicit analytic expression for
$\Gamma_{2}$ can then be written down in general, but it is not especially
illuminating, although it can be very useful for numerical calculations. It
simplifies greatly for $q\mapsto0$, where we get simply: $C(q=0)=(|\omega
|/2c)\sqrt{\omega^{2}-4\Delta_{0}^{2}}$.

The (integrable) square-root singularity at the edges of the branch cuts is
present at finite $q$ as well, and indeed, for $\omega^{2}\gtrsim2\Delta
_{0}^{2}(q)$, we obtain, to leading order: $C(q)\approx\alpha(q)\sqrt
{\omega^{2}-4\Delta_{0}^{2}(q)}+\mathcal{O}(\omega^{2}-4\Delta_{0}^{2}(q))$
with: $\alpha(q)=(\Delta_{0}(q)/2c\Delta_{0})\{|\omega|-q^{2}c^{2}/2\Delta_{0}(q)\}.$

By exploiting the parity of $\Gamma_{2}, \Gamma_{1}$ will be given
then, via dispersion relations, by:
\begin{equation}
\Gamma_{1}(q,\omega)=2\int\limits_{2\Delta_{0}(q)}^{\infty}\frac
{d\omega^{\prime}}{\pi}\omega^{\prime}
\Gamma_{2}(q,\omega^{\prime})\mathcal{P}(\frac{1}{\omega^{\prime2}-\omega^{2}})
\end{equation}
which shows that $\Gamma_{1}(q,\omega)$ will be strictly \textbf{positive} for $|\omega|<2\Delta_{0}(q)$, a result that will prove to be useful shortly.

Let's consider now the full longitudinal propagator:%
\begin{equation}
\widetilde{G}_{c}^{L}(q,\Omega_{n})=gcS^{2}\frac{(3\widetilde{\Gamma}%
(q,\Omega_{n})/2gc)}{(3\widetilde{\Gamma}(q,\Omega_{n})/2gc)[\Omega_{n}%
^{2}+\varepsilon^{2}(q)]+(m_{s}/S)^{2}}%
\end{equation}

Performing the analytic continuation, omitting specification of the label
$q$ and defining:%
\begin{equation}
G(z)=\widetilde{G}_{c}^{L}(q,\Omega_{n})/gcS^{2},\text{ \ }\Gamma
(z)=3\widetilde{\Gamma}(q,\Omega_{n})/2gc,\text{ \ }\delta=(m_{s}%
/S)^{2},\text{ \ }\epsilon=\varepsilon(q)
\end{equation}
we are led to study the analytic structure of a function of the form:%
\begin{equation}
G(z)=\frac{\Gamma(z)}{\Gamma(z)(\epsilon^{2}-z^{2})+\delta}%
\end{equation}
($0\leq\delta<1,$ $\epsilon<2\Delta_{0}$). $\Gamma(z)$ will be given by:%
\begin{equation}
\Gamma(z)=\int\frac{d\omega^{\prime}}{\pi}\frac{\Gamma_{2}(\omega^{\prime}%
)}{\omega^{\prime}-z}%
\end{equation}
with $\Gamma_{2}(\omega)$ having all the properties that have been listed
above (odd in $\omega$, positive for positive $\omega$ and vanishing for
$|\omega|\leq2\Delta_{0}$ (cfr. Eq.($61$)), thus producing a branch-cut in
$\Gamma(z)$ for real $z=\omega$ and $2\Delta_{0}\leq|\omega|<+\infty$). We
will write: $\Gamma(\omega+i0^{+})=\Gamma_{1}(\omega)+\Gamma_{2}(\omega)$ \ on
the real axis. The analytic properties of $G(z)$ will be determined in turn
by its spectral weight function.

Going to the real axis we find, for $z=\omega+i\eta,\eta>0$:%
\begin{equation}
G(\omega+i\eta)=G_{1}(\omega)+iG_{2}(\omega)
\end{equation}
where, defining:%
\begin{equation}
A(\omega)=(\epsilon^{2}-\omega^{2}+\eta^{2})\Gamma_{1}(\omega)+2\eta
\omega\Gamma_{2}(\omega)+\delta
\end{equation}%
\begin{equation}
B(\omega)=2\eta\omega\Gamma_{1}(\omega)-(\epsilon^{2}-\omega^{2}+\eta
^{2})\Gamma_{2}(\omega)
\end{equation}
$G_{1}$and $G_{2}$ are given by:%
\begin{equation}
G_{1}(\omega)=\frac{A(\omega)\Gamma_{1}(\omega)-B(\omega)\Gamma_{2}(\omega
)}{A^{2}(\omega)+B^{2}(\omega)}%
\end{equation}
and:%
\begin{equation}
G_{2}(\omega)=\frac{A(\omega)\Gamma_{2}(\omega)+B(\omega)\Gamma_{1}(\omega
)}{A^{2}(\omega)+B^{2}(\omega)}%
\end{equation}

Even for $\eta\mapsto0$, $G_{2}$ needs not vanish when $\Gamma_{2}$ does. In
particular, let's inspect its structure for $|\omega|<2\Delta_{0}.$ Sending
$\eta$ to $0$ inside the $\Gamma$'s (and only there) we have, for
$|\omega|<2\Delta_{0}$: $\ A(\omega)=(\epsilon^{2}-\omega^{2}+\eta^{2}%
)\Gamma_{1}(\omega)+\delta,$ $B(\omega)=2\eta\omega\Gamma_{1}(\omega)$,
leading to:%
\begin{equation}
G_{1}(\omega)=\frac{\epsilon^{2}-\omega^{2}+\eta^{2}+\delta/\Gamma_{1}%
(\omega)}{(\epsilon^{2}-\omega^{2}+\eta^{2}+\delta/\Gamma_{1}(\omega
))^{2}+4\eta^{2}\omega^{2}}%
\end{equation}
and:%
\begin{equation}
G_{2}(\omega)=\frac{2\eta\omega}{(\epsilon^{2}-\omega^{2}+\eta^{2}%
+\delta/\Gamma_{1}(\omega))^{2}+4\eta^{2}\omega^{2}}%
\end{equation}

From the structure of $G_{2}(\omega)$ it is clear that:%
\begin{equation}
\underset{\eta\mapsto0}{\lim}G_{2}(\omega)=\pi sgn(\omega)\delta
(f(\omega));\text{ }f(\omega)=\omega^{2}-\epsilon^{2}-\frac{\delta}{\Gamma
_{1}(\omega)}%
\end{equation}

Remembering that $\Gamma_{1}$ is an even function of $\omega$, we can write:
$\Gamma_{1}=\Gamma_{1}(\omega^{2}),$ and $f(\omega)=0$ will have solutions
at $\omega=\pm\epsilon_{L}$ with:
\begin{equation}
\epsilon_{L}^{2}=\epsilon^{2}+\frac{\delta}{\Gamma_{1}(\epsilon_{L}^{2})}%
\end{equation}
that reduces (as it should) to $\epsilon^{2}=\varepsilon^{2}(q)$ when
$\delta\mapsto0$. Moreover:%
\begin{equation}
\frac{df}{d\omega}=2\omega\frac{df}{d\omega^{2}}=2\omega\left[1+\frac{\delta
}{\Gamma_{1}^{2}(\omega^{2})}\frac{d\Gamma_{1}(\omega^{2})}{d\omega^{2}}\right]
\end{equation}
\ According to known formulas, then:%
\begin{equation}
\underset{\eta\mapsto0}{\lim}G_{2}(\omega)=\gamma\frac{\pi}{2\epsilon_{L}%
}\{\delta(\omega-\epsilon_{L})-\delta(\omega+\epsilon_{L})\}
\end{equation}
where:%
\begin{equation}
\gamma=\left\{\left[1+\frac{\delta}{\Gamma_{1}^{2}(\omega^{2})}\frac{d\Gamma_{1}
(\omega^{2})}{d\omega^{2}}\right]_{\omega=\epsilon_{L}}\right\}^{-1}%
\end{equation}
This proves of course that the longitudinal propagator has (in the range we
are examining) simple poles on the real axis at: $\omega=\pm\epsilon_{L}$, and
the prefactor $\gamma$ will give the reduction of the quasiparticle weight
w.r.t. the pure bosonic case.

\begin{figure}[t]
\begin{center}
\hspace*{.1cm}
\epsfxsize=7cm \epsffile{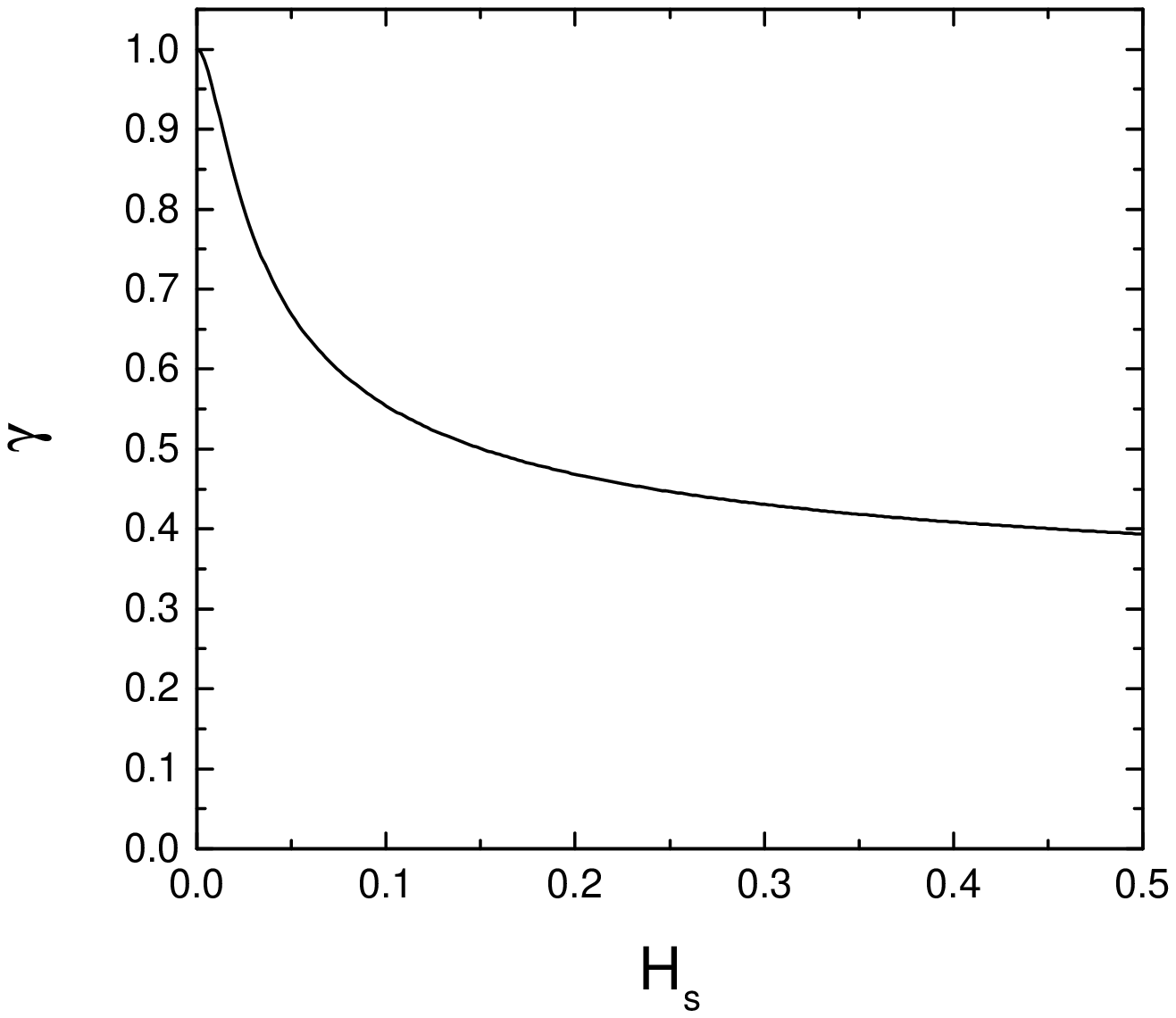}
\vspace{.4cm}
\caption[]{\footnotesize 
The relative quasiparticle weight of the pole of the longitudinal
propagator as a function of the staggered field and at $q=0$.}
\end{center}
\label{fig2}
\end{figure}
As, for small fields, $\delta\propto H_{s}^{2}$, $\gamma$ will approach $1$
quadratically in the field when $H_{s}\mapsto0$. A numerical plot of the
relative quasiparticle weight $\gamma$ at $q=0$ is presented in Fig. $2$. It
is a steadily decreasing function of the field, and the quadratic regime near
$H_{s}=0$ is confined to a very narrow region of fields. For higher fields
there is an intermediate region in which $\gamma$ is almost linear in the
field, and we find numerically that for $S=1$ it saturates in the high-field
limit to:%
\begin{equation}
\underset{H_{s}\mapsto\infty}{\lim}\gamma=\underset{\delta\mapsto1}{\lim
}\gamma\simeq0.279
\end{equation}

More than $70\%$ of the quasiparticle weight is then lost when the system
evolves towards saturation. No significant changes are expected for $q\neq0$.

To complete the analysis we have to investigate the range $|\omega
|>2\Delta_{0}$ of $G_{2}(\omega)$ where $\Gamma_{2}(\omega)$ does not vanish.
With some long but straightforward algebra we find:%
\begin{equation}
G_{2}(\omega)=\frac{\delta\Gamma_{2}(\omega)}{(\epsilon^{2}-\omega^{2}%
)^{2}(\Gamma_{2}(\omega))^{2}+(\delta+(\epsilon^{2}-\omega^{2})\Gamma
_{1}(\omega))^{2}}%
\end{equation}
for $|\omega|>2\Delta_{0}$. Therefore, in this range of frequencies $G_{2}$
will vanish as $H_{s}\mapsto0$ and we will recover \ the simple pole structure
with the longitudinal and transverse propagators becoming equal. 
The longitudinal pole will survive
also up to saturation, but with a strongly field-dependent strength.

That as the field increases the spectral weight that is lost from the pole
gets transferred to the two-magnon continuum $(80)$ (and viceversa when the
field decreases) is dictated, e.g., by the sum rule:
\begin{equation}
\int\limits_{-\infty}^{+\infty}\frac{d\omega}{\pi}\omega\operatorname{Im}%
G_{2}(\omega)=1
\end{equation}
The sum rule $(81)$ is just one of the general sum rules connected with
the moment expansions of the spectral weight functions that are related to
equal-time expectation values of multiple commutators and that have been known
for a long time in many-body theory \cite{16,17}. A proof of the sum rule adapted to the specific context of the NL$\sigma$M will be given in the
Appendix\footnote{Note however that while (cfr. Eq.$(52)$) all moments exist
for the transverse propagator, when $G_{2}$ is given by Eqs.$(77)$ and $(80)$
only the first moment will exist and all the others will turn out to be
divergent. This is just an artifact of the mean-field approximation 
(see, e.g., the discussion of a similar problem in Ref.\cite{17}.}.

The pole at $\epsilon_{L}$ represents the longitudinal magnon. It will be a
well defined excitation as long as $\epsilon_{L}^{2}(q)<4\Delta_{0}^{2}(q)$,
which we will prove to be the case. It will be higher in energy (as we have
proved previously that $\Gamma_{1}(q,\omega)>0$ for $|\omega|<2\Delta_{0}(q)$)
than the two (degenerate) transverse magnons that both have energy $\epsilon$, and will become degenerate with the latter when $\delta\mapsto0$ (the limit
in which the full $SO(3)$ invariance is restored).

Just as in the transverse case ($\epsilon=\varepsilon(q)=\sqrt{c^{2}%
q^{2}+\Delta_{T}^{2}}$, $\Delta_{T}=c\xi^{-1}=\Delta_{0}$), we can define a
longitudinal gap $\Delta_{L\text{ }}$via: $\Delta_{L\text{ }}=\epsilon
_{L}(q=0)$, i.e.:
\begin{equation}
\Delta_{L}^{2}=\Delta_{T}^{2}+\frac{\delta}{\Gamma_{1}(0,\Delta_{L})}%
\end{equation}

In general:%
\begin{equation}
\Gamma_{1}(0,\omega)=\frac{g^{2}c}{2}{\underset{\eta\mapsto0}\lim}%
\int\limits_{2\Delta_{0}(q)}^{+\infty}\frac{d\omega^{\prime}}{\pi}\frac
{1}{\sqrt{(\omega^{\prime2}-4\Delta_{0}^{2})}}\frac{\omega^{\prime2}%
-\omega^{2}}{(\omega^{\prime2}+\omega^{2}+\eta^{2})^{2}-4\omega^{2}%
\omega^{\prime2}}%
\end{equation}
As we are assuming here: $\Delta_{L}<2\Delta_{T}$, the limiting procedure
becomes trivial and we obtain:%
\begin{equation}
\Gamma_{1}(0,\Delta_{L})=\frac{3}{4}g\int\limits_{2\Delta_{T}}^{+\infty}%
\frac{d\omega}{\pi}\frac{1}{\sqrt{(\omega^{2}-4\Delta_{T}^{2})}}\frac
{1}{\omega^{2}-\Delta_{L}^{2}}%
\end{equation}

Eq.($82$) can be rewritten in dimensionless form as:%
\begin{equation}
\left(\frac{\Delta_{L}}{\Delta_{T}}\right)^{2}=1+\frac{4\delta
}{3gF(\Delta_{L}/\Delta_{T})}%
\end{equation}
where:%
\begin{equation}
F(y)=\int\limits_{2}^{+\infty}\frac{dx}{\pi}\frac{1}{\sqrt{x^{2}-4}}\frac
{1}{x^{2}-y^{2}};\text{ \ }|y|<2
\end{equation}
Explicitly:%
\begin{equation}
F(y)=\frac{1}{2\pi y\sqrt{1-y^{2}/4}}\left\{\frac{\pi}{2}-\tan^{-1}%
\left[\frac{2}{y}\sqrt{1-y^{2}/4}\right]\right\}
\end{equation}
and, defining: $y=\Delta_{L}/\Delta_{T}$ we obtain, with: $g=2/S$:
\begin{equation}
y^{2}=1+\frac{8m_{s}^{2}yS\sqrt{1-y^{2}/4}}{3\left\{1-\frac{2}{\pi}%
\tan^{-1}\left[\frac{2}{y}\sqrt{1-y^{2}/4}\right]\right\}}
\end{equation}

\begin{figure}[t]
\begin{center}
\hspace*{.1cm}
\epsfxsize=7cm \epsffile{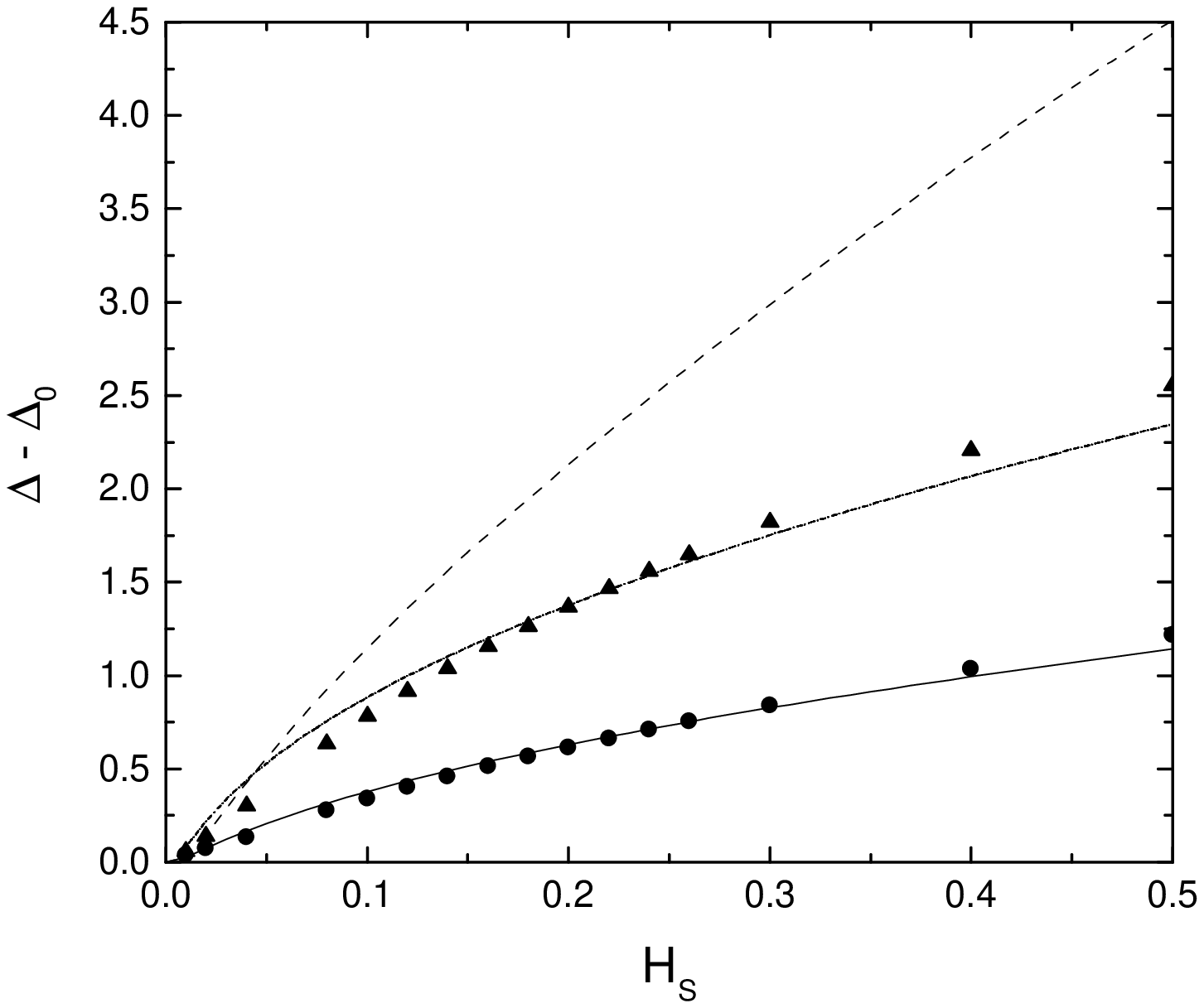}
\vspace{.4cm}
\caption[]{\footnotesize 
Results for the spin gaps in the longitudinal and transverse channels:\\
-Dots and full line: DMRG results of Ref.\cite{6} and our mean-field 
results for the transverse gap.\\
-Triangles and dot-dashed line: DMRG results of Ref.\cite{6} and our mean-field
results for the longitudinal gap.\\
-Dashed line: SMA predictions for the longitudinal gap (Ref.\cite{5} and our
Ref.\cite{7}).}
\end{center}
\label{fig1}
\end{figure}

The numerical results for the longitudinal gap are reported for $S=1$ in
Fig. 3, where we report also the results for the transverse gap. Here too
the results are in excellent qualitative agreement with the DMRG results,
with some small quantitative discrepancies in the low-field regime. We also
find that near $H_{s}=0$ the longitudinal gap increases roughly as three times
the transverse one (actually: $\lim_{H_{s}\mapsto0}(\Delta_{L}-\Delta
_{0})/(\Delta_{T}-\Delta_{0})\simeq3.58$). For $H_{s}\mapsto\infty$ instead:%
\begin{equation}
y^{2}=1+\frac{8yS\sqrt{1-y^{2}/4}}{3\left\{1-\frac{2}{\pi}\tan^{-1}%
\left[\frac{2}{y}\sqrt{1-y^{2}/4}\right]\right\}}
\end{equation}
and, for $S=1$:%
\begin{equation}
\underset{H_{s}\mapsto\infty}{\lim}\frac{\Delta_{L}}{\Delta_{T}}\simeq1.855
\end{equation}
i.e. for high fields $\Delta_{L}$ tends to increase slightly less than twice
$\Delta_{T}$. This is in agreement with the DMRG results of Ref.\cite{6}. 
Notice that, however, the limiting form of Eq.$(89)$ tells us immediately that the ratio of the saturation values of the gaps will tend exactly to two in the
large-$S$ \ limit, and indeed it is not difficult to see that:
\begin{equation}
\underset{H_{s}\mapsto\infty}{\lim}\frac{\Delta_{L}}{\Delta_{T}}\approx
2-\frac{a^{2}}{S^{2}}
\end{equation}
for $S\gg1$, with $a$ a numerical constant of order $0.5$. The relative
quasiparticle weight $\gamma$ can also be shown to be of order $S^{-2}$ in the
same limit, i.e. it will vanish when the magnon poles reaches the edge of the continuum.

\section{Discussion and conclusions}

\bigskip

We summarize here our results, comparing them at the same time with those
obtained within other approaches. We will list and discuss only a few relevant
points, namely:

$i)$ In zero field the excitation spectrum consists of the well-known
degenerate triplet of massive Haldane bosons with energy $\varepsilon
(q)=c\sqrt{q^{2}+\xi^{-2}}$ with a gap $\Delta_{0}=c\xi^{-1}.$ For finite
fields, instead:

$ii)$ The staggered magnetization curve (Eq.$(29)$ and Fig.$1$) turns out to
agree well with the DMRG results of Ref.\cite{6}. As can be deduced directly from Eq. $(29)$ and as was discussed in more detail in Ref.\cite{7}, the low- and high-field behaviours of the staggered magnetization are respectively:
\begin{equation}
m_{s}\approx\chi^{T}H_{s}+\mathcal{O}(H_{s}^{3})
\end{equation}
for $H_{s}\approx0$ and:%
\begin{equation}
m_{s}\approx S\left[1-\frac{A}{\sqrt{H_{s}}}+\mathcal{O}(H_{s}^{-1})\right]
\end{equation}
with $A$ a numerical constant (see \cite{7} for more details) for large $H_{s}$.
The value obtained in \cite{7} of $\chi^{T}=23.74/J$ is somewhat higher than the
DMRG result \cite{6} of $\chi^{T}=18.5/J$. We will resume this point shortly below.

Eq.$(93)$ shows that the staggered magnetization saturates only
asymptotically in the large-field limit. This is what had to be expected, of
course, and is a direct consequence of our implementing in a consistent way
the NL$\sigma$M constraint. The authors of Ref.\cite{5} found instead a
magnetization curve that (see Fig.$1$) agrees better with the DMRG data than
ours in the initial part (i.e. for low fieds), but that disagrees more and
more as the field is increased. Even worse, their magnetization saturates at a
finite value of the staggered fields and keeps growing indefinitely beyond
saturation afterwards. Full saturation implies of course that the fully
polarized N\'{e}el state should become the exact ground state of the
Hamiltonian $(1)$, which known to be true only asymptotically for
$H_{s}\mapsto\infty$. Moreover, an $\mathit{over}$saturated magnetization is
clearly not an admissible result.

It appears therefore that softening the NL$\sigma$M constraint as done in
Ref.\cite{5} can be an admissible procedure in the zero-field limit, that can be also given (see \cite{18} and references therein) some legitimation in the same
limit, but that can become more and more dangerous as the field increases,
leading eventually to rather unphysical results.

$iii)$ The transverse channel is saturated by a single well-defined
magnon pole with energy $\epsilon_{T}(q)=\varepsilon(q)$ and a gap $\Delta
_{T}=\Delta_{0}.$The strength of the pole has a rather weak field dependence
and the pole persists up to the highest fields. The relation: $\chi
_{T}=Sgc/(\Delta_{T})^{2}$, which is typical of the SMA, is exactly obeyed
in this channel at the mean-field level. 
Both our results and those of Ref.\cite{5} agree quite well with the DMRG results.

$iv)$ In the longitudinal channel instead we find also a well defined magnon
pole with energy $\epsilon_{L}(q)$ and a gap $\Delta_{L}>\Delta_{T}$, but the
magnon propagator acquires also a (two-magnon) branch-cut for nonzero
staggered fields. The strength of the pole of the propagator at $\epsilon_{L}$
is strongly field-dependent and as the field increases it is steadily
transferred to the continuum, in agreement with the sum rule $(81)$. Near
saturation, and for $S=1$, the strength of the pole is reduced to less than
$30\%$ of its zero-field value, while it vanishes completely at saturation in
the large-$S$ limit. In the same limit the longitudinal pole disappears into
the continuum at saturation, while it remains below the continuum (hence a
well-defined excitation, although with a strongly reduced intensity) for
finite $S$. The relation analogous to case $iii)$, namely: $\chi
_{L}=Sgc/(\Delta_{L})^{2}$ is badly violated in this case (cfr. Fig.$3$)
\ except in the zero-field limit, and this proves that, due to a non
negligible contribution from the two-magnon continuum, the SMA is not
applicable at all in the longitudinal channel in the presence of a
nonvanishing (staggered) field, as was done in \cite{5} (see the close critical
comparison with the DMRG results that was made in Ref.\cite{6}, and the SMA
curve that is reported for comparison in Fig.$3$). This makes a great
difference with the zero-field case, where there is quite convincing evidence
\cite{19} that two- and/or multi-magnon excitations carry a negligible spectral
weight, which turns out to be actually exactly zero at the mean-field level.

In conclusion, we would like to point to two somewhat unsatisfactory
aspects of the theoretical scenario that has been outlined here and elsewhere
\cite{5}-\cite{7}.

First of all, and as we have already remarked, there is excellent qualitative
and even semiquantitative agreement between our NL$\sigma$M analysis and the
DMRG results over the whole range of fields. There remain however some
quantitative discrepancies. As to these, we can only stress the fact that the
NL$\sigma$M mapping is basically a semiclassical expansion starting from the
large-$S$ \ limit that is then continued to lower values of the spin. Being
so, and if it has to have any validity at all, it should definitely be able to
capture all the essential features of the low-energy physics of models such as
that of Eq.$(1)$, which we believe we have proved to be the case for $S=1$. It
should then lead definitely to qualitative and semiquantitative agreement, but
it cannot be expected to yield also, and it would be actually a totally
unexpected bonus if it did, precise quantitative agreement with, e.g., the
DMRG results. We believe that, apart from other refinements \cite{13}, not much more than that can be expected from the NL$\sigma$M mapping.

In view of what has been just said, however, the quantitative agreement should
improve for increasing values of the spin. A test of the NL$\sigma$M, in
conjunction with a DMRG analysis, on higher-spin chains can be of some
interest \cite{20}.

Coming now to the experimental scenario outlined in the Introduction, while
essentially all the theoretical models presented so far (including ours) yield
quite reasonable agreement between theory and experiment as far as the
staggered magnetization curve and the transverse gap are concerned, recent
measurements \cite{4} of the longitudinal gap in Nd$_{2}$BaNiO$_{5}$ point to some discrepancies between theory and experiment. 
Namely, the longitudinal gap is
found to survive, with $\Delta_{L}>\Delta_{T}$, also above the N\'{e}el
temperature, i.e. also when $H_{s}=0$, which points to (single-ion)
anisotropies that, although could be accomodated very easily within the
NL$\sigma$M approach, thus leading to an additional spin-gap in the
longitudinal channel, are not explicitly accounted for by the models
considered so far. Moreover, below $T_{N}$, the longitudinal gap appears to
grow more slowly than the transverse one, and this remains a so far unresolved puzzle.

More measurements on compunds of the class $R_{2}$BaNiO$_{5}$ are called for to
ascertain whether this is a general fact or it is specific to 
Nd$_{2}$BaNiO$_{5}$. 
If the former were the case, this would imply that a model
involving just a single chain in an effective staggered field, interesting in
itself and worth of theoretical interest in its own, although it can account
for most of the experimental observations, is not enough to provide a complete
description of all the observed low-energy-physics features of such compounds,
and that more sophisticated models are needed.

\bigskip

\bigskip
 
\bigskip

\section{Appendix. A sum rule for the NL$\sigma$M\\ propagators}

To cope with the notation adopted in Sect.$4$ after Eq.$(64)$, let's consider
(setting: $g=c=S=1$), in real time now, a classical Lagrangian of the form:%

\begin{equation}
\mathcal{L}(x,t)=\frac{1}{2}\left\{|\partial_{t}\mathbf{n}|^{2}-|\partial
_{x}\mathbf{n}|^{2}-V(\mathbf{n})\right\}
\end{equation}
with $V(\mathbf{n})$ an arbitrary interaction term, e.g. one implementing in
some appropriate limit \cite{15,16} the NL$\sigma$M constraint 
$\mathbf{n}^{2}(x,t)=1$. 
The canonical momentum density is defined by:
\begin{equation}
\mathbf{\pi}(x,t)=\frac{\partial\mathcal{L}}{\partial(\partial_{t}\mathbf{n}%
)}=\partial_{t}\mathbf{n}%
\end{equation}
and the Hamiltonian density will be given, as usual, by: $\mathcal{H}%
(x,t)=\mathbf{\pi\cdot}\partial_{t}\mathbf{n}-\mathcal{L},$ with canonical
(equal-time) Poisson brackets (PB's):
\begin{equation}
\{n^{\alpha}(x,t),\pi^{\beta}(y,t)\}=\delta^{\alpha\beta}\delta(x-y)
\end{equation}
with all the other PB's vanishing.

Canonical quantization (here too we set: $\hbar=1$ and use for simplicity the
same notation for the classical variables and the corresponding quantum
operators) is accomplished by replacing the nonvanishing PB's with
the commutators:
\begin{equation}
\lbrack n^{\alpha}(x,t),\pi^{\beta}(y,t)]=i\delta^{\alpha\beta}\delta(x-y)
\end{equation}

Then, we obtain immediately, taking expectation values in the ground state:%
\begin{equation}
-i\partial_{t}\langle\lbrack n^{\alpha}(x,t),n^{\beta}(0,0)]\rangle
|_{t=0}=-i\langle\lbrack\pi^{\alpha}(x,0),n^{\beta}(0,0)]\rangle=-\delta(x)
\end{equation}
As the r.h.s. involves only equal-time commutators, this relation will be true
irrespective of the form of the interaction term as long as the latter depends
only on $\mathbf{n}$ and not on its time derivative.

Introducing the Fourier transform $A^{\alpha\beta}(k,\omega)$ defined via:%
\begin{equation}
\langle\lbrack n^{\alpha}(x,t),n^{\beta}(0,0)]\rangle=\int\frac{dk}{2\pi}%
\int\frac{d\omega}{2\pi}\exp\{i[kx-\omega t]\}A^{\alpha\beta}(k,\omega)
\end{equation}
Eq.$(98)$ will imply the sum rule:%
\begin{equation}
\int\frac{d\omega}{2\pi}\omega A^{\alpha\beta}(k,\omega)=\delta^{\alpha\beta}%
\end{equation}
Let's test it in the case: $V(\mathbf{n})=m^{2}\mathbf{n}^{2}$, with $m$
fixed, e.g., by the condition: $\langle\mathbf{n}^{2}(x,t)\rangle=1$. Then the
field can be quantized \cite{18} as\footnote{Note that, in order to reproduce correctly the commutation relations (97)
between the fields and the conjugate momenta, the signs of the arguments of
the exponentials had to be taken here as the opposites of those of Ref.\cite{18}.}:
\begin{equation}
n^{\alpha}(x,t)=\int\frac{dk}{4\pi\varepsilon(k)}\left\{a_{k}%
\exp[i(kx-\varepsilon(k)t)]+a_{k}^{\dagger}\exp[-i(kx-\varepsilon
(k)t)]\right\}
\end{equation}
($\varepsilon(k)=\sqrt{k^{2}+m^{2}}$), and:%
\begin{equation}
\lbrack n^{\alpha}(x,t),\pi^{\beta}(y,t)]=i\delta^{\alpha\beta}\delta
(x-y)\Leftrightarrow\lbrack a_{k}^{\alpha},a_{q}^{\beta\dagger}]=4\pi
\delta^{\alpha\beta}\varepsilon(k)\delta(k-q)
\end{equation}
One finds then easily:
\begin{equation}
\langle\lbrack n^{\alpha}(x,t),n^{\beta}(0,0)]\rangle=\delta^{\alpha\beta}%
\int\frac{dk}{2\pi}\exp(ikx)\frac{1}{2\varepsilon(k)}[\exp(-i\varepsilon
(k)t)-\exp(i\varepsilon(k)t)]
\end{equation}
whence:%
\begin{equation}
A^{\alpha\beta}(k,\omega)=\delta^{\alpha\beta}
\frac{\pi}{\varepsilon(k)}\{\delta(\omega-\varepsilon(k))-
\delta(\omega+\varepsilon(k))\}
\end{equation}
and the sum rule can be checked at once.

To compare with Eq.$(81)$, let's recall first of all that our definition
(Eq.$(17)$) of the Euclidean propagators differs by a sign from that that is
usually adopted \cite{21}. Accordingly, we will define the retarded Green
functions of the components of the $\mathbf{n}$ field as:%
\begin{equation}
G_{R}^{\alpha\beta}(x,t)=i\theta(t)\langle\lbrack n^{\alpha}(x,t),n^{\beta
}(0,0)]\rangle
\end{equation}
and their Fourier transform are then given by:
\begin{equation}
\widetilde{G}_{R}^{\alpha\beta}(k,\omega)=-\int\frac{d\omega^{\prime}}{2\pi
}\frac{A^{\alpha\beta}(k,\omega^{\prime})}{\omega-\omega^{\prime}+i\delta
}|_{\delta\mapsto0^{+}}%
\end{equation}
Hence:%
\begin{equation}
\operatorname{Im}\widetilde{G}_{R}^{\alpha\beta}(k,\omega)=\frac{1}%
{2}A^{\alpha\beta}(k,\omega^{\prime})
\end{equation}
In view of the standard relationship \cite{21,22} between retarded (in real time) and causal (in Euclidean time) Green functions, the sum rule that we have just derived extends to the Euclidean propagators introduced in the text and
becomes then the sum rule $(81)$.

\vspace{2cm}

\hrule

\end{document}